# Ruddlesden-Popper faults in LaNiO$_3$/LaAlO$_3$ superlattices

E. Detemple,[1] Q. M. Ramasse,[2] W. Sigle,[1] G. Cristiani,[3] H.-U. Habermeier,[3] B. Keimer,[3] P. A. van Aken[1]

[1] Max Planck Institute for Intelligent Systems, 70569 Stuttgart, Germany

[2] SuperSTEM Laboratory, STFC Daresbury, WA4 4AD Warrington, United Kingdom

[3] Max Planck Institute for Solid State Research, 70569 Stuttgart, Germany

ABSTRACT

Scanning transmission electron microscopy in combination with electron energy-loss spectroscopy is used to study LaNiO$_3$/LaAlO$_3$ superlattices grown on (La,Sr)AlO$_4$ with varying single-layer thicknesses which are known to control their electronic properties. The microstructure of the films is investigated on the atomic level and the role of observed defects is discussed in the context of the different properties. Two types of Ruddlesden-Popper faults are found which are either two or three dimensional. The common planar Ruddlesden-Popper fault is induced by steps on the substrate surface. In contrast, the three-dimensionally arranged Ruddlesden-Popper fault, whose size is in the nanometer range, is caused by the formation of local stacking faults during film growth. Furthermore, the interfaces of the superlattices are found to show different sharpness, but the microstructure does not depend substantially on the single-layer thickness.

INTRODUCTION

Recent research has shown that oxide heterostructures and superlattices (SLs) show fascinating phenomena which are absent in their bulk constituents and can result in qualitatively different electronic properties.[1] In addition, they can be systematically controlled through parameters such as epitaxial strain,[2] the electronic dimensionality,[3] or by external fields[4] which makes them interesting candidates for new electronic devices.[5] Nevertheless, as in the case of bulk materials, the macroscopic device properties can be critically affected by the microstructure of the entire film, e.g. in the form of dislocations[6-8] or extended planar faults[9] which are commonly (and often inadvertently) generated during fabrication and can strongly affect the transport of charge carriers through the device.[8] Therefore, the defect structure has to be investigated if the properties are correlated with the above mentioned parameters (strain, dimensionality, external fields).

Among the variety of transition metal oxides, $LaNiO_3$ (LNO) is a promising representative for heterostructuring due to its strongly correlated conduction electrons. Bulk LNO behaves as a correlated metal over the complete temperature range, but a metal–insulator transition was reported as soon as the LNO conduction electrons are confined in SLs of atomically thin LNO layers between insulating $LaAlO_3$ (LAO) blocking layers.[3] The properties of SLs with thicker LNO layers match those of bulk LNO. Related transport phenomena were also reported in ultrathin LNO films.[10] Theoretical studies even predict superconductivity in suitably prepared LNO SLs.[11,12]

$LaNiO_3/LaAlO_3$ (LNO/LAO) SLs can be grown with high precision on different substrates such as $SrTiO_3$ or $(La,Sr)AlO_4$ (LSAO) as is demonstrated exemplarily by atomically resolved elemental maps in Fig. 1 for a SL grown on $SrTiO_3$. Here, we report about a detailed characterization of $LaNiO_3/LaAlO_3$ (LNO/LAO) SLs grown on LSAO with different single-layer thicknesses by means of transmission electron microscopy (TEM) in conventional as well as in the scanning mode in combination with electron energy-loss spectroscopy (EELS). The high quality of the SL is demonstrated and it is shown that the abruptness of the two interfaces (LAO–LNO and LNO–LAO) is different. Nevertheless, the films contain defects, namely Ruddlesden-Popper-type faults (RP faults) which are known to occur in perovskite thin films.[13-17] In our films, we observe two different configurations of these RP faults whose origin and exact atomic arrangements are described in detail.

EXPERIMENTAL

The LNO/LAO SLs are epitaxially grown by pulsed laser deposition on (La,Sr)AlO$_4$ (LSAO) substrates using a KrF excimer laser with 2 Hz pulse rate and 1.6 J/cm$^2$ energy density. A compressive strain is induced by LSAO due to its smaller lattice parameter in comparison to LNO and LAO. After the deposition in 0.5 mbar oxygen atmosphere at 730 °C, the films are subsequently annealed in 1 bar oxygen atmosphere at 690 °C for 30 min. Since the properties of these SLs, e.g. the conductivity, depend on the thickness of the individual layers,[3] SLs containing single layers of two or four unit cells thickness were investigated in order to see how the microstructure is affected by the single-layer thickness.

For the TEM studies, the samples were mechanically thinned by tripod polishing. Additionally, some samples, cooled with liquid nitrogen, were shortly ion milled using low ion accelerating voltages (0.5–1 kV). A $C_s$-corrected Nion UltraSTEM operated at 100 kV acceleration voltage was used to record high-angle annular dark field (HAADF) images and both, atomically resolved EELS linescans and maps. The convergence angle of the electron beam was 31 mrad while the collection angle of the spectrometer was 32 mrad. Multivariate statistical analysis was performed to reduce the noise of the EEL spectra.[18] High-resolution transmission electron microscopy (HRTEM) images were recorded with a JEOL 4000FX microscope operated at 400 kV acceleration voltage.

RESULTS AND DISCUSSIONS

As the HAADF image in Fig. 2(a) exemplarily shows, the single layers of all our LNO/LAO SLs are epitaxially grown and very homogeneous at a large scale which demonstrates the high quality of the SLs. The area within the white rectangle is enlarged in Fig. 2(b) resolving the lanthanum, nickel, as well as the aluminum atomic columns. Since the intensity is approximately proportional to $Z^{1.7}$, the brightest spots correspond to lanthanum which is visible all over the layer system. The weaker spots in between show nickel and aluminum atom columns whose intensities differ significantly because of the Z difference. This allows distinguishing LNO from LAO layers and shows that the single layers are well defined. Obviously, the bottom layer in Fig. 2(b) is LAO, followed by LNO and so on. An EELS linescan was recorded in the growth direction from position A (bottom) to B (top) and is depicted in Fig. 2(b). The linescan is horizontally integrated over the width of Fig. 2(b) to improve the signal-to-noise ratio. The intensity of the La $M_5$, Ni $L_2$, and Al K edges were extracted after background subtraction. The Ni $L_2$ edge is chosen instead of the Ni $L_3$ edge because of its reduced overlap with the La $M_4$ edge. The profiles of the Ni $L_2$ and Al K intensities are plotted in Fig. 2(c) confirming the clear separation of the single layers of the SL. However, a closer look at the aluminum profile [red line in Fig. 1(c)] shows an asymmetric shape. The left side is very steep whereas an additional peak is present on the right side at the position of the first $NiO_2$ plane of the LNO layer (marked with black arrows). Therefore the abruptness of the two interfaces is different, a feature that has also been reported for other perovskite SLs.[19-21] These results show that the LAO–LNO interface is rougher than the LNO–LAO interface in LNO/LAO SLs.

Contrary to LNO/LAO SLs on non-polar SrTiO$_3$ substrates which show NiO precipitates,[22] no formation of secondary phases was observed in the present films grown on polar LSAO substrate. This is shown in the HRTEM image of the entire film (Fig. 3). However, extended planar defects perpendicular to the layers are visible. Their separation shows strong spatial variations and ranges from 5 to 100 nm. The planar faults are present in all samples irrespective of the single-layer thickness. The HAADF image of the interface between the LSAO substrate and the SL reveals a surface step of the substrate underneath the planar defect. This surface step appears to be the origin of the planar defect [Fig. 4(a)]. According to the literature, the height of the surface step is 4.534 Å[23] which differs from the LNO lattice parameter (3.838 Å).[24] The termination of LSAO is different on both sides of the step, as shown by the overlaid LSAO unit cells. On the left, the substrate is terminated by a (La,Sr)O plane so that the LNO growth starts with a NiO$_2$ plane. By contrast, on the right side of the surface step, an AlO$_2$ plane is on top and consequently an LaO plane is first deposited. Taking into account the lattice mismatch, the lattices on both sides of the defect are vertically shifted against each other. This becomes especially evident in the elemental EELS maps of Fig. 4 (b) and (c). At the fault, La atoms on one side face Ni or Al atoms on the other side. In addition to this displacement, the vertical NiO$_2$ plane along the planar fault is missing [see upper arrow in Fig. 4 (a)] resulting in a zigzag arrangement of the lanthanum atoms along the defect. The complete defect structure can also be described by two lattices which are displaced by a displacement vector of 1/2 [111]. This type of defect is known as RP fault whose structure is similar to the RP phases.[13] RP faults typically develop in nonstoichiometric films with an excess of the larger cation.[14-16] We conclude, however, the occurrence of the RP faults in our films is not caused by insufficient control of the stoichiometry during the growth process as they were not found in films which were grown under the same conditions on other substrates like SrTiO$_3$.[22] Rather, we believe that the real source for the development of the RP faults are the surface steps of the LSAO substrate since every imaged RP fault can be traced to such a step. This is confirmed by the fact that the density of the RP faults can be directly correlated to the number of surface steps on the substrate. At the same time, we note that the impact of the observed RP faults on the stoichiometry is negligible because their density is very low. Concerning a possible strain relaxation of the film, the missing plane along the RP fault could in principle allow the film to relax part of the compressive strain which is induced by the LSAO substrate. However, it is hard to decide in how far the RP faults really cause a relaxation because their density varies strongly.

The blocks marked by dashed rectangles in the HAADF image shown in Fig. 5 (a) show a second defect type. The enlarged image of a single block [Fig. 5 (b)] shows that all atomic columns have similar brightness within the block. This does not imply that the atoms of the block have the same $Z$ because TEM images show a projection of the crystal. The homogeneous contrast means that within the atom columns the average atomic number is similar. The reason is the existence of cuboid-shaped blocks with limited size which are displaced by 1/2 [111] with respect to the host lattice. The situation is sketched in the 3D atomic model [Fig. 5 (c)] which exactly represents the marked area of Fig. 5 (b), neglecting oxygen atoms. The viewing direction is along the black arrow, i.e. the atomic columns are either pure lanthanum, pure nickel, or pure aluminum in the case of the perfect SL. This results in the varying intensity of the dots in the non-defective areas. By contrast, if one follows the atomic columns along the viewing direction in the upper left part of the model, which corresponds to a faulted cuboid, the columns emerge as mixed resulting in a similar intensity of all columns. The zigzag arrangement of the lanthanum atoms along the borders of the block shows that RP faults terminate the blocks in all three dimensions (3D). In the following, we will hence refer to this defect as a 3D RP fault.

Although both defect structures can be described as RP faults, they differ significantly in some points, e.g. in their size. A lower limit of the dimension of the planar faults must be the thickness of the TEM samples (25–40 nm) and thus it can be classified as a 2D defect. On the other hand, the 3D RP fault is a comparatively small 3D inclusion surrounded by RP faults. The length of the cuboid edges parallel to the substrate is only a few nanometers and the height is about 15 nm. A further difference between the two types is their origin. The planar RP faults are directly correlated with the substrate because they are the consequence of the surface steps of the substrate. In the case of the 3D RP fault, no correlation with the substrate could be found but rather they are likely to be a product of small environmental variations during the growth process. We suggest the following growth model which is illustrated in Fig. 6. Lanthanum atoms are represented by blue, nickel atoms by yellow and aluminum by red balls; the oxygen atoms are neglected. We start with a flat LNO layer which is terminated by an LaO plane [Fig. 6 (a)]. In the case of perfect growth, the subsequent plane would be $AlO_2$ followed by LaO and so on [left side of Fig. 6 (b) and (c)]. A 3D RP fault forms when a further LaO plane is locally deposited instead of $AlO_2$ on top of the final LaO plane of the LNO layer [right side of Fig. 6 (b)]. This local stacking fault is embedded in the normally grown $AlO_2$ plane and causes the formation of a 3D RP fault. Above this plane, regular growth resumes, so that LaO is deposited on top of $AlO_2$ and vice versa as Fig. 6 (c) shows. The 3D RP fault grows until a further stacking fault terminates it. After further planes are deposited, the cross-sections of both cases look as plotted in Fig. 6 (d). On the left side, the layered structure of the perfect SL is visible. The right cross-section illustrates that the proposed deposition sequence really results in such blocks whose borders are RP faults which are visible as the zigzag arrangement of the lanthanum atoms.

The 3D RP faults occur irrespective of the thickness of the individual layers of the SL. However, a higher density of 3D RP faults can be expected in SLs with thinner single layers, because the number of interfaces is higher for the same total thickness of the film. Altogether, the microstructure of the films is not affected by the single-layer thickness so that differences in the phase behavior between LNO/LAO SLs with two or four unit cell thick single layers cannot be caused by the microstructure of the films but rather reflect the confinement of the conduction electrons in the thinner single layers.[3]

CONCLUSIONS

In summary, the atomic structure of LNO/LAO SLs was studied by scanning TEM and EELS measurements. It was found that the LNO–LAO interface is more abrupt than the LAO–LNO interface. Furthermore, two types of RP faults differing in their dimension have been detected. The 2D RP fault is an extended planar defect which originates at surface steps of the LSAO substrate. In contrast, the formation of small cuboids which are bordered by RP faults is induced by local stacking faults which can develop during the deposition of the films. Their growth nucleation is not related to the substrate. The microstructure of the SLs has turned out to be independent of the single-layer thickness of the SLs. Consequently, the microstructure is not the determining factor concerning the differences of properties between SLs with two or four unit cell thick single layers but rather the differences are an inherent phenomenon of the SL.

FIGURES

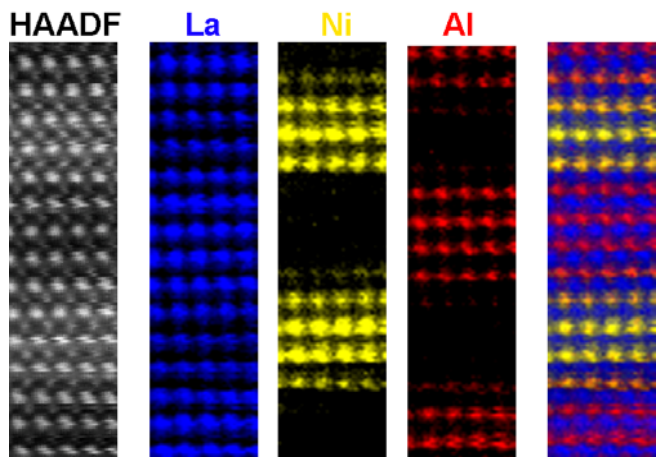

Fig. 1: HAADF image (left) and the respective elemental EELS maps of a LNO/LAO superlattice with 4 u.c. thick single layers grown on SrTiO$_3$ substrate. The right image is an overlay of the La, Ni, and Al maps. The horizontal image width is 1.8 nm. The aluminium and nickel atoms are arranged in layers while the lanthanum pattern is continuous.

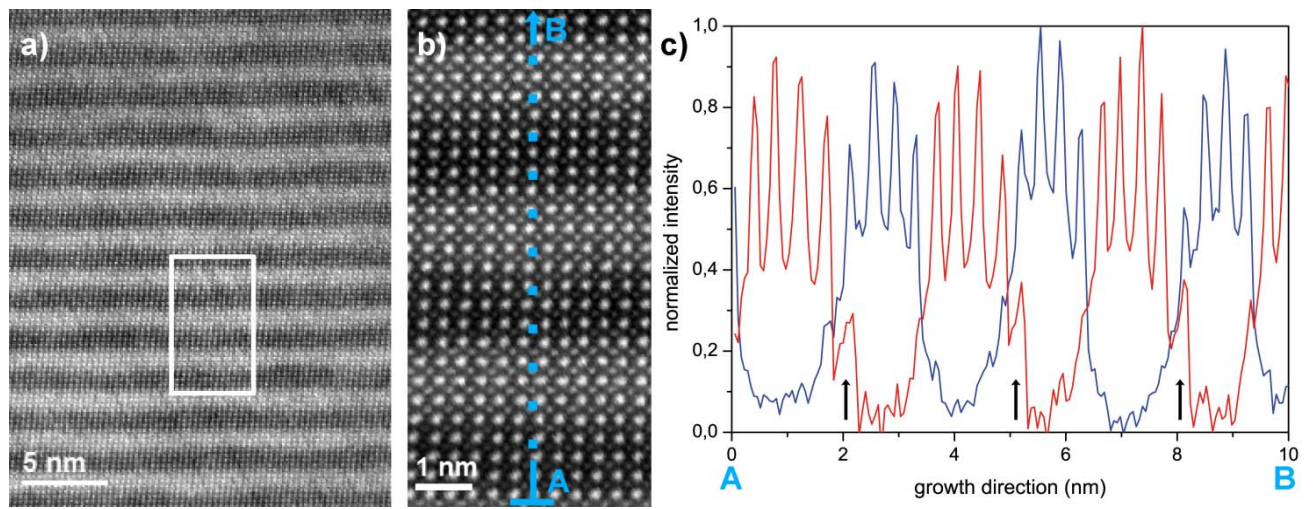

Fig. 2: HAADF images of a LNO/LAO superlattice with 4 u.c. thick single layers: a) Overview b) white rectangular in a) at higher magnification. The brightest spots are the La columns, in between weaker Ni and Al columns are visible. C) Integrated EELS linescan over the horizontal width of image b) from the bottom (A) to the top (B). The profiles of the normalized intensities of the Ni L$_2$ (blue) and Al K (red) edges are plotted.

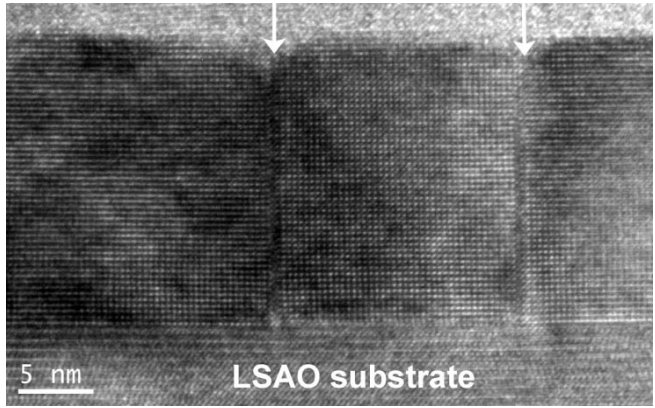

Fig. 3: HRTEM image of the planar RP faults which are marked by arrows.

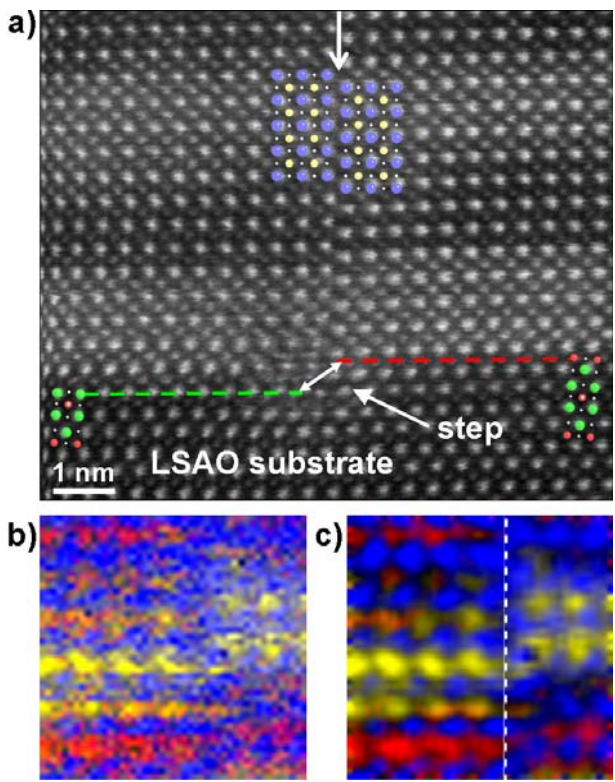

Fig. 4: a) HAADF image of a planar RP fault (marked by the upper arrow) near the interface to the substrate. The unit cells of LSAO and LNO are overlaid: La blue, Ni yellow, O grey, (La,Sr) green, and Al red. The termination of the substrate differs on the two sides of the RP fault. On the left, the substrate is terminated by a (La,Sr)O plane, on the right by a $AlO_2$ plane. This results in a surface step of the substrate which is the origin of the RP fault. The unit cells of LNO show that the atomic planes from the both sides are shifted against each other and that one $NiO_2$ plane is missing. Elemental EELS maps (2.56 x 2.56 nm²) are shown in b) and c) (low pass filtered). The dashed line in c) shows the position of the RP fault.

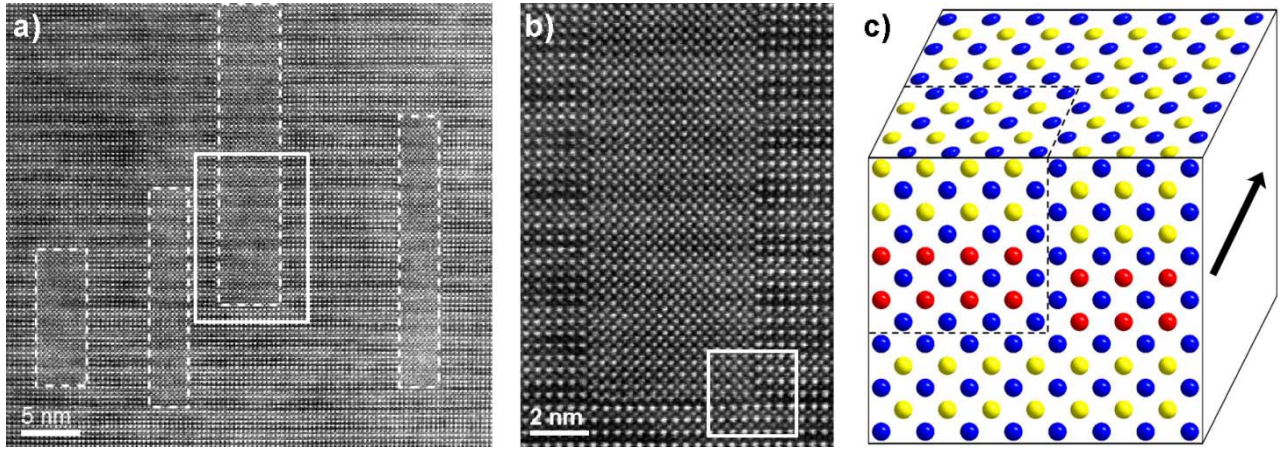

Fig. 5: a) HAADF image showing several blocks (marked by dotted rectangles). b) Enlarged image from the area within the solid rectangle in a) showing a single block which consists of columns with similar brightness. c) 3D atomic model of the white rectangle in b) La blue, Ni yellow, and Al red. The blocks are surrounded by RP faults (marked by dotted lines). In the perfect superlattice, there are pure La, Ni, and Al columns along the viewing direction (black arrow). In contrary, the columns are mixed in the upper left part resulting in the similar intensity of all columns within the block.

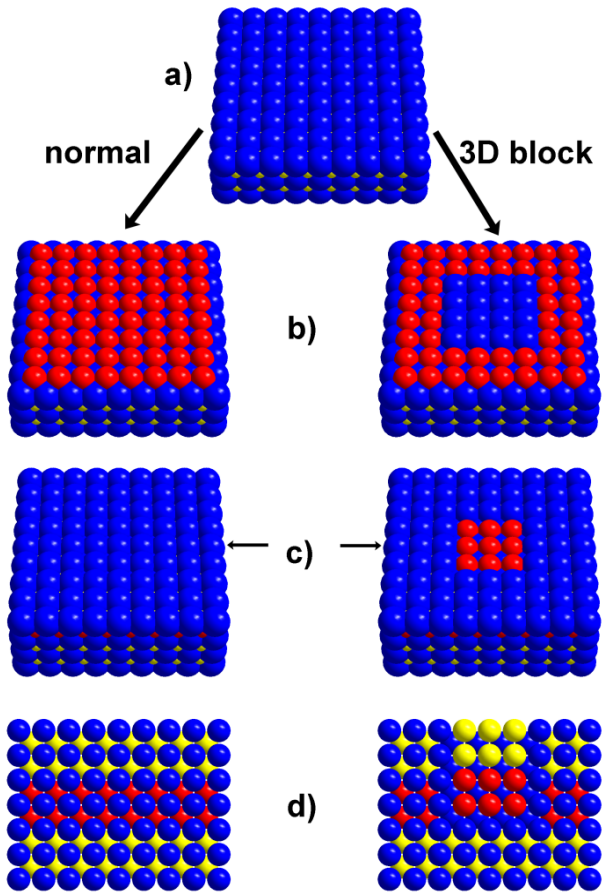

Fig. 6: Atomic model showing the normal growth of a superlattice on the left side and the growth of a 3D RP fault on the right. La blue, Al red, and Ni yellow; O is not shown for simplicity. The starting point is in both cases a perfectly flat LaO plane (a). On top of it an AlO$_2$ plane is deposited in the case of perfect growth, in contrary LaO is locally deposited (surrounded by AlO$_2$) in the case of the growth of a 3D RP fault (b). This is followed by normal growth: LaO on top of AlO$_2$ and AlO$_2$ on top of LaO (c). The cross-sections whose position is marked by the arrow in c) show the perfect superlattice and the 3D block after further deposition steps (d).